Note on "Unsteady MHD flow due to non-coaxial rotations of a porous disk and a fluid at infinity", by T. Hayat et al., Acta Mech. 151, 127-134 (2001).

Tarek M. A. El-Mistikawy
Department of Engineering Mathematics and Physics, Faculty of Engineering, Cairo University, Giza 12211, Egypt

**Abstract**

The cited article contains a serious flaw. A remedy is found in articles by other authors.

**Analysis**

The authors of Ref. [1] formulated the following problem for the unsteady MHD flow due to non-coaxial rotations of a fluid at infinity and a porous disk.

$$\nabla \cdot V = 0 \qquad (1)$$

$$\frac{\partial V}{\partial t} + (V \cdot \nabla)V = -\frac{1}{\rho}\nabla p + \nu \nabla^2 V - \frac{\sigma}{\rho}B_0^2 V \qquad (2)$$

where V is the velocity vector whose components $(u, v, w)$ in the Cartesian coordinate directions $(x, y, z)$ satisfy the boundary and initial conditions

$$u = -\Omega y, \quad v = \Omega x, \quad w = -W_0 \quad \text{at} \quad z = 0 \quad \text{for} \quad t > 0 \qquad (3\text{a-c})$$

$$u = -\Omega(y - l), \quad v = \Omega x \quad \text{as} \quad z \to \infty \quad \text{for} \quad t \geq 0 \qquad (4\text{a,b})$$

$$u = -\Omega(y - l), \quad v = \Omega x \quad \text{at} \quad t = 0 \quad \text{for} \quad z \geq 0 \qquad (5\text{a,b})$$

All symbols are defined in Ref. [1] except for $W_0$ which, obviously, represents a suction velocity. The right-most term in Eq. (2) represents the electromagnetic body force, simplified under the assumption of small magnetic Reynolds number and the unrealistic choice of an applied magnetic field of strength $B_0$ that is perpendicular to the unknown velocity vector V.

Following Erdogan [2], who handled the same problem in the absence of the magnetic field, the authors assumed the velocity components to have the forms

$$u = -\Omega y + f(z, t), \quad v = \Omega x + g(z, t), \quad w = -W_0 \qquad (6\text{a-c})$$

which satisfy Eq. (1) and give for Eq. (2)

$$\frac{1}{\rho}\frac{\partial p}{\partial x} = \Omega^2 x + \Omega g + W_0 \frac{\partial f}{\partial z} - \frac{\partial f}{\partial t} + \nu \frac{\partial^2 f}{\partial z^2} - \frac{\sigma}{\rho}B_0^2(f - \Omega y) \qquad (7\text{a})$$

$$\frac{1}{\rho}\frac{\partial p}{\partial y} = \Omega^2 y - \Omega f + W_0 \frac{\partial g}{\partial z} - \frac{\partial g}{\partial t} + \nu \frac{\partial^2 g}{\partial z^2} - \frac{\sigma}{\rho} B_0^2 (g + \Omega x) \qquad (7b)$$

$$\frac{1}{\rho}\frac{\partial p}{\partial z} = \frac{\sigma}{\rho} B_0^2 W_0 \qquad (7c)$$

The article [1] has a serious flaw. Differentiating Eq. (7a) with respect to $y$ and Eq. (7b) with respect to $x$, leads to

$$\frac{\partial^2 p}{\partial x \partial y} = +\sigma B_0^2 \Omega \qquad (8a)$$

$$\frac{\partial^2 p}{\partial y \partial x} = -\sigma B_0^2 \Omega \qquad (8b)$$

which violate the continuity of the pressure $p$. The assumed forms (6a-c) for the velocity components are incompatible with the model in use.

An alternative model that allows for the forms (6a-c) to be applicable is proposed in Refs. [3,4]. It involves an electric field $E = -B_0 \Omega(x, y - l, 0)$, and invokes the more realistic magnetic field $B = B_0(0,0,1)$ in place of the unrealistic one used in Ref. [1].

For comparison with Eqs. (7a-c), the alternative model gives, instead

$$\frac{1}{\rho}\frac{\partial p}{\partial x} = \Omega^2 x + \Omega g + W_0 \frac{\partial f}{\partial z} - \frac{\partial f}{\partial t} + \nu \frac{\partial^2 f}{\partial z^2} - \frac{\sigma}{\rho} B_0^2 (f - \Omega l) \qquad (9a)$$

$$\frac{1}{\rho}\frac{\partial p}{\partial y} = \Omega^2 y - \Omega f + W_0 \frac{\partial g}{\partial z} - \frac{\partial g}{\partial t} + \nu \frac{\partial^2 g}{\partial z^2} - \frac{\sigma}{\rho} B_0^2 g \qquad (9b)$$

$$\frac{1}{\rho}\frac{\partial p}{\partial z} = 0 \qquad (9c)$$

where, now

$$\frac{\partial^2 p}{\partial x \partial y} = \frac{\partial^2 p}{\partial y \partial x} = 0 \qquad (10a)$$

sustaining the continuity of $p$, and

$$\frac{\partial^2 p}{\partial x^2} + \frac{\partial^2 p}{\partial y^2} = 2\rho \Omega^2 \qquad (10b)$$

which integrates to $p = p_0 + \frac{1}{2}\rho \Omega^2 [x^2 + (l - y)^2]$, leading upon substitution in Eqs. (9a,b) to

$$\frac{\partial f}{\partial t} - W_0 \frac{\partial f}{\partial z} - \Omega g = \nu \frac{\partial^2 f}{\partial z^2} + \frac{\sigma}{\rho} B_0^2 (\Omega l - f) \tag{11a}$$

$$\frac{\partial g}{\partial t} - W_0 \frac{\partial g}{\partial z} - \Omega(\Omega l - f) = \nu \frac{\partial^2 g}{\partial z^2} - \frac{\sigma}{\rho} B_0^2 g \tag{11b}$$

With $F = f + ig$, $i = \sqrt{-1}$, Eqs. (11) combine to give

$$\nu \frac{\partial^2 F}{\partial z^2} + W_0 \frac{\partial F}{\partial z} - \frac{\partial F}{\partial t} + \left(i\Omega + \frac{\sigma}{\rho} B_0^2\right)(\Omega l - F) = 0 \tag{12}$$

subject to the conditions

$$F(0, t) = 0, F(\infty, t) = \Omega l, F(z, 0) = \Omega l \tag{13}$$

This problem agrees with Eqs. (2.11) and (2.13) of Ref. [1], whose solution is given therein.

**Note 1**: In Ref. [5], a more general analysis of the problem is performed, indicating the need for an electric field $\mathrm{E} = [E_x(x, y, t), E_y(x, y, t), 0]$ which satisfies $\nabla \cdot \mathrm{E} = -2B_0\Omega$. The electric field $\mathrm{E} = -B_0\Omega(x, y - l, 0)$ used in Refs. [3] and [4] is, obviously, a special case.

**Note 2**:, The flaw described in this note is found in other publications, among which are Ref. [6]-[19].

**Note 3**: References [17]-[19] replace the MHD flow by flow through a porous medium, with similar expression for the body force. In this case no remedy is possible.

**References**

[1] Hayat, T., Asghar, S., Siddiqui, A. M., Haroon, T.: Unsteady MHD flow due to non-coaxial rotations of a porous disk and a fluid at infinity. Acta Mech. **151**, 127-134 (2001).

[2] Erdogan, M. E.: Unsteady flow of a viscous fluid due to non-coaxial rotations of a disk and a fluid at infinity, Int. J. Non-Linear Mech. **12**, 285-290 (1997).

[3] Chakraborti, A., Gupta, A. S., Das, B. K., Jana, R. N.: Hydromagnetic flow past a rotating porous plate in a conducting fluid rotating about a noncoincident parallel axis, Acta Mech. **176**, 107–119 (2005).

[4] Guria, M., Das, S., Jana, R.N.: Hall effects on unsteady flow of a viscous fluid due to non-coaxial rotation of a porous disk and a fluid at infinity, International Journal of Non-Linear Mechanics **42**, 1204 – 1209 (2007).

[5] Ersoy, H. Volkan, MHD flow of an Oldroyd-B fluid due to non-coaxial rotations of a porous disk and the fluid at infinity, International Journal of Engineering Science **38**, 1837-1850 (2000).